\documentstyle[11pt]{article}
\begin{document}
\title{Charged Higgs boson pair production via gluon-gluon collisions in MSSM
       with CP violation
\footnote{This work was supported by National Natural Science Foundation of
          China.}}

\vspace{-3mm}
\author{{Jiang Yi$^{b}$, Ma Wen-Gan$^{a,b}$, Han Liang$^{b}$, Hou Hong-Sheng$^{b}$}\\
{\small $^{a}$CCAST (World Laboratory), P.O.Box 8730, Beijing 100080, China} \\
{\small $^{b}$Modern Physics Department, University of Science and}\\
{\small Technology of China, Anhui 230027, China} }

\date{}
\maketitle
\vspace{-8mm}

\begin{abstract}
The CP-violating effects to the subprocess $gg\rightarrow H^+H^-$ is studied
in the mSUGRA scenario at the CERN LHC, by taking into account the
experimental bounds of electron and neutron EDM's. The CP violation effects
in this process are related to the complex phases of $\mu$ and $A_{f}$ in
the mSUGRA scenario. In our calculation we consider small CP phases of
$\mu$ and $A_{f}$ and neglect the effects of neutral Higgs bosons mixing.
In this case the CP effects to the process mainly come from the complex
couplings of Higgs-squark-squark. We find a strong dependence of
charged Higgs boson pair production rate on the complex couplings
in the MSSM parameter space.

\end{abstract}
{\large\bf PACS: 14.80.Cp, 12.60.Jv,12.15.Ji }\\

\noindent{\large\bf Key words: Charged Higgs Boson, mSUGRA, CP Violation}

\vfill \eject


\noindent{\large\bf I. Introduction}

\par
One of the basic questions of particle physics is the nature of
electroweak symmetry breaking. Various theoretical insights suggest the
extensions from the minimal Higgs sector with one Higgs doublet, for example,
the Minimal Supersymmetric Standard Model (MSSM). The MSSM predicts the
existence of three neutral and two charged Higgs bosons $h^0$, $H^0$,
$A^0$ and $H^{\pm}$\cite{MSSM}, in which two Higgs doublets must be
introduced to provide masses to down- and up-type fermions through the
superpotential and to keep the theory free of anomalies. The lightest neutral
Higgs boson in the MSSM may be difficult to distinguish from the neutral
Higgs boson of the standard model(SM). The exploration of additional Higgs
bosons will be then very important to confirm extended Higgs sector.
Discovery of a charged Higgs boson, $H^{\pm}$, shows directly these extended
versions of the Higgs sector. Beyond the discovery of Higgs bosons,
reconstructing the Higgs potential will be necessary, and that requires the
experimental study of the self-couplings of Higgs bosons.
\par
At the Large Hadron Collider(LHC), searching for these extra Higgs bosons
is also one of the most important tasks. In much of the parameter space
preferred by the MSSM, namely $m_{H^{\pm}}>m_{W}$ and $1<\tan\beta<m_{t}/m_{b}$
\cite{pspace}, the LHC will provide several mechanisms to search for charged
Higgs boson. A copious source of charged Higgs bosons with fair light mass
are decay of top quarks $t\rightarrow bH^{+}$\cite{tdecay}, since top quarks
are produced with very large rates at the LHC. For a heavier charged Higgs
boson the dominate production process is $gb\rightarrow tH^{-}$\cite{gbth}.
The pair charged Higgs bosons can be produced via quark-antiquark annihilation
\cite{qaqa} and gluon-gluon fusion\cite{ pairjy, pairk, pairb, pairh}. Of course, Higgs pair
production has a lower rate compared with the single production mechanism, but
Higgs pair production via $gg$ and $q\bar{q}$ annihilation involves the
trilinear and quartic couplings of the physical Higgs bosons. So the
measurement of these couplings will be useful in reconstructing the
Higgs potential.
\par
The full MSSM Lagrangian consists of various soft masses and the
higgsino $\mu$ parameter and has 45 CP-violating complex phases\cite{cpphase}.
In principle, the CP phases may come from the soft SUSY-breaking gaugino masses
$M_{a}$ $(a=U(1)_{Y}, SU(2)_{L}, SU(3)_{C})$, the soft bilinear Higgs mass
mixing masses $m^{2}_{12}$ ( or $B_{\mu}$), the soft trilinear Yukawa
couplings $A_{f}$ and the higgsino mass parameter $\mu$. Assuming universality
condition of the soft gaugino masses at the GUT (or Planck) scale, the gaugino
masses have a common phase and the different soft trilinear Yukawa coupling
are all equal at this scale. Due to the two global $U(1)$ symmetries of the
MSSM Lagrangian, the $U(1)_{Q}$ Peccei-Quinn symmetry and the $U(1)_{R}$
symmetry acting on the Grassman-valued coordinates, not all CP phases are
physical and the physical CP phases could be reduced to those of trilinear
coupling $A_{f}$ and the $\mu$ parameter\cite{gsy}, $\Phi_{\mu}$ and
$\Phi_{A_{f}}$,
$$
\mu=|\mu|e^{i\Phi_{\mu}},~~~A_{f}=|A_{f}|e^{i\Phi_{A_{f}}}.
\eqno{(1.1)}
$$
The CP phases are constrained indirectly by the measurements of the
fermionic electric dipole moment(EDM), mainly of electron EDM $d_{e}$
and neutron EDM $d_{n}$. In order to satisfy the current experimental
bounds of electron\cite{de} and neutron \cite{dn} EDMs, there are several
possibilities to solve it. The first possibility is to let the CP phases
very small, i.e.$|\Phi|\leq 10^{-2}$ \cite{smallcp}. The second possibility
is that the CP phases are not very small but the first two generations of
scalar fermions which contribute to EDMs are heavy\cite{heavy}. The third
possibility is that the partial cancellations among various contributions
to the electron and neutron EDM's in a restricted parameter space and the
CP phases do not have to be suppressed\cite{cancellation1, can2}.
\par
In the Higgs sector of the MSSM, at the tree-level Higgs potential is
CP invariant, since the mass parameters of the two Higgs fields and the
vacuum expectation values $\tan\beta$ are real and $\mu$ enters only via
$|\mu|^2$.  If one includes the radiative corrections involving
soft-CP-violating Yukawa interactions in the theory, the tree-level CP
invariance of the Higgs potential can be broken\cite{hcp1, hcp2}. It's so
called explicit CP violation of the MSSM Higgs potential. Then the three
neutral Higgs bosons can mix and none of the mass eigenstate Higgs boson is
a CP eigenstate. For $|\mu|$ and $|A_{f}|$ values larger than the arithmetic
average of the scalar-top-quark masses squared, denoted as $M_{SUSY}$($M^{2}
_{SUSY}=1/2(m^{2}_{\tilde{t}_{1}}+ m^{2}_{\tilde{t}_{2}})$), the explicit
CP-violating effects can be significant\cite{hcp1}. If $|\mu|$ and $|A_{f}|$
values are smaller than $M^{2}_{SUSY}$ or the CP phases of $\mu$ and $A_{f}$ is
much small, such CP-violating effects turn out to be negligible.
\par
Various SUSY processes including the effects of the CP phases have
recently been considered, such as the effects in the neutral Higgs boson
production at hadron collider\cite{dedes}, the effects in neutralino
\cite{neu} and chargino \cite{cha} production at LEP, at CERN LHC\cite{mazhou},
as well as at future electron-positron linear colliders\cite{lc}. In this work
we shall investigate the CP violation effects in process $ g g \rightarrow
H^{+}H^{-}$ in the context of the mSUGRA. In the numerical calculation we
choose $|\mu|$ and $|A_{f}|$ smaller than $M^{2}_{SUSY}$ and the CP phases of
$\mu$ and $A_{f}$ at the order $O(10^{-2}-10^{-1})$ to neglect the effects of
the neutral Higgs bosons mixing. While the constrains to the parameters from
the measurements of the electron and neutron EDM's are taken into account. We
organized the paper as follows.  In Section II, the analytical calculations
are presented.  In section III there are numerical results and discussions and
in section IV, a short summary is given.

\vskip 5mm
\noindent{\large\bf II. Calculation}

The Feynman diagrams contributing to the subprocess $gg\rightarrow H^+H^-$
and the total cross section of this subprocess can be found in \cite{pairjy}.
The effective Lagrangian for electric dipole interaction of a fermion
can be written as
$$
{\cal L}=-\frac{i}{2}d_{f}\bar{\psi}\sigma_{\mu\nu}\gamma_{5}\psi F^{\mu\nu}.
\eqno{(2.1)}
$$
At the one-loop level, the contribution in the framework of MSSM to the
electron EDM comes from the virtual chargino and neutralino exchanges, for the
neutron EDM it is contributed by the virtual chargino, neutralino and
gluino exchange. By considering the
electric dipole operator, the chromoelectric dipole operator and the purely
gluonic six-dimension operator, one found \cite{cancellation1} that there
exist significant regions in the parameter space where cancellations between
the gluino and the chargino exchanges reduce the electric and the
chromoelectric contributions, and the purely gluonic parts lead to a dramatic
lowering of the neutron EDM sometimes below the electron EDM value. In
the case of the neutron EDM, the dominant chargino and gluino contributions
appear with opposite sign over a large portion of the MSSM parameter space.
In general, internal cancellations are more likely among the SUSY contributions
to the neutron EDM than they are in the case of the electron. A recent
calculation of Chang et al.\cite{chang} for the Barr-Zee-type\cite{barr}
two-loop contributions to the EDMs, shows that the sbottom contribution is
enhanced for a large $\tan\beta$\cite{baer}. Therefore, in our analysis we
choose suitable parameters and small CP phases to suppress the Barr-Zee-type
two-loop contributions. The CKM CP phase in the SM can in principle
contribute to the fermionic EDM, such as the electron EDM, but it
turns out to be effective at three-loop level so that the contribution is
very small.
\par
The mSUGRA is based on $N=1$ supergravity grand unified theory, in
which supersymmetry can be broken spontaneously via gravitational
interactions and the electroweak symmetry is broken via radiative effect
\cite{radiat}. At the GUT scale, the CP-violating mSUGRA can be completely
parameterized by just six input parameters, namely, $m_{1/2}$, $m_{0}$,
$A_{0}$, $\tan\beta$, $\Phi_{\mu}$ and $\Phi_{A_{0}}$, where $m_{1/2}$,
$m_{0}$ and $A_{0}$ are the universal gaugino mass, scalar mass and the
trilinear soft breaking parameter, $\Phi_{\mu}$ and $\Phi_{A_{0}}$ are the
CP phases of $\mu$ and the trilinear soft breaking parameter. From these
six parameters, all the masses and coupling of the model are determined by
the evolution from the GUT scale to the low electroweak scale.
\par
With the cross section of the subprocess, we can easily obtain the
total cross section at hadron collider by folding the cross section of
$gg\rightarrow H^+H^-$ with the gluon luminosity
$$
\sigma(pp \rightarrow gg+X \rightarrow
          H^{+}H^{-}+X)=
    \int_{4 m^2_{H^+}/s }^{1}
d\tau \frac{d{\cal L}_{gg}}{d\tau} \hat{\sigma}(gg \rightarrow
          H^{+}H^{-},
     \hskip 3mm {\rm at} \hskip 3mm \hat{s}=\tau s). \eqno{(2.2)}
$$
where $\sqrt{s}$ and $\sqrt{\hat{s}}$  denote the proton-proton
and gluon-gluon c.m.s. energies respectively and
$\frac{d{\cal L}_{gg}} {d\tau}$ is the gluon luminosity, which is defined as
$$
\frac{d{\cal L}_{gg}}{d\tau}=\int_{\tau}^{1}
 \frac{dx_1}{x_1} \left[ f_{g}(x_1,Q^2) f_{g}(\frac{\tau}{x_1},Q^2) \right].
 \eqno{(2.3)}
$$
Here we used $\tau=x_{1}x_{2}$, the definitions of $x_1$ and $x_2$ are adopted
from Ref.\cite{jiang}. In our numerical calculation we take the MRST(mode 2)
parton distribution function $f_{g}(x_{i}, Q^{2})$ \cite{MRST}, and ignore
the supersymmetric QCD corrections to the parton distribution functions for
simplicity. The factorization scale $Q$ was chosen as the average of the
final particle masses $m_{H^{+}}$. The numerical calculation is carried out
for the LHC at the energy $14 TeV$.

\vskip 10mm
\noindent
{\Large{\bf III. Numerical Results and Discussions}}
\vskip 5mm
In this section, we present some numerical results of the CP effects to
total cross section from the full one-loop diagrams involving virtual
(s)quarks for the process $pp \rightarrow gg +X\rightarrow H^{+}H^{-} +X$.
The input parameters are chosen as: $m_t=173.8~GeV$, $m_{Z}=91.187~GeV$,
$m_b=4.5~GeV$, $\sin^2{\theta_{W}}=0.2315$, and $\alpha = 1/128$.
We adopt a simple one-loop formula for the running strong coupling
constant $\alpha_s$
$$
\alpha_s(\mu)=\frac{\alpha_{s}(m_Z)} {1+\frac{33-2 n_f} {6 \pi} \alpha_{s}
              (m_Z) \ln \left( \frac{\mu}{m_Z} \right) }. \eqno{(3.1)}
$$
where $\alpha_s(m_Z)=0.117$ and $n_f$ is the number of active flavors at
energy scale $\mu$.
In our numerical calculation to get the low energy scenario from the mSUGRA,
the renormalization group equations(RGE's)\cite{RGE} are run from
the weak scale $M_Z$ up to the GUT scale, taking all threshold into account.
We use two loop RGE's only for the gauge couplings and the one-loop
RGE's for the other supersymmetric parameters. The GUT scale boundary
conditions are imposed and the RGE's are run back to $M_Z$, again taking
threshold into account. The REG's for the gauge and Yukawa couplings, and
the diagonal elements of the sfermion masses and gaugino masses are real.
The phase of $\mu$ does not run because it cancels out in the one loop
renormalization group equation of $\mu$, and the magnitude and the
phase of $A_{f}$ do evolve\cite{cancellation1}. The decay widths of the
intermediate Higgs bosons are considered at the tree level and these formula
can be found in ref.\cite{hunter}.
\par
In order to describe the CP violating effects in the process, we
introduce the notation $\delta$ defined as
$$
\delta=\frac{\sigma_{CP}-\sigma_0}{\sigma_0},
\eqno{(3.1)}
$$
where $\sigma_0$ is the cross section of this process with CP conservation
and $\sigma_{CP}$ is the cross section with CP violations. In Fig.1 we present
the $\delta$ versus CP phases of $\mu$ and $A_{0}$. The input parameters are
chosen as $m_{0}=100 ~GeV$,$m_{1/2}=150~GeV$, $A_{0}=300 ~GeV$, $\tan\beta=4$
and the CP phases of $\mu$ and $A_{0}$ run from zero to $0.1\pi$. When the CP
phases are set zero, the masses of Higgs bosons are $m_{h^0}=97 ~GeV$,
$m_{H^0}=258 ~GeV$, $m_{A^0}=254 ~GeV$ and $m_{H^+}=266 ~GeV$ and the cross
section of this process is $1.2 ~fb$. In our numerical calculation we let
$\Phi_{\mu}=0$ when $\Phi_{A_{0}}$ runs from zero to $0.1 \pi$ and vice versa.
In this case, the CP phase corrections to the masses of Higgs bosons and squarks are the order
of just a few percent. So the CP effects to the process at this case mainly
come from the phases in the Higgs-squark-squark couplings. The CP effect of
$\Phi_{\mu}$ and $\Phi_{A_{0}}$ to this process increase as the phase becoming
larger. The $\delta$ is positive for $\Phi_{\mu}$ running in this region
and can reach $12\%$
when the $\Phi_{\mu}=0.1\pi$, but for $\Phi_{A_{0}}$, the $\delta$ is negative
and can reach $-46\%$ when the $\Phi_{A_{0}}=0.1\pi$. From the figure we
can see that the effect of $\Phi_{A_{0}}$ is larger than that of $\Phi_{\mu}$ to
this process. There is a strong dependence of the production rates of the pair
charged Higgs on the complex couplings in such MSSM parameter space.
\par
In Fig.2 we also present the $\delta$ versus CP phases of $\mu$ and
$A_{0}$. The input parameters are chosen as $\tan\beta=30$ and the others
are the same as those in Fig.1. When the CP phases are set zero, the masses
of Higgs bosons are $m_{h^0}=106 ~GeV$, $m_{H^0}=175 ~GeV$, $m_{A^0}=174 ~GeV$
and $m_{H^+}=194 ~GeV$ and the cross section of this process is $44 ~fb$.
We still let $\Phi_{\mu}=0$ when $\Phi_{A_{0}}$ runs from zero to $0.1 \pi$
and vice versa. In this case, the CP phases corrections to the masses of Higgs bosons and
squarks are very small and the CP effects to the process mainly
come from the CP phases in the Higgs-squark-squark couplings. The CP effects of
$\Phi_{\mu}$ and $\Phi_{A_{0}}$ to this process increase as the phase becoming
larger. Both the $\delta$ of $\Phi_{\mu}$ and $\Phi_{A_{0}}$ to this process
are negative.  For $\Phi_{\mu}$ the $\delta$ can reach $-5.7\%$ when the
$\Phi_{\mu}=0.1\pi$ and for $\Phi_{A_{f}}$ can reach $-12\%$ when the
$\Phi_{A_{0}}=0.1\pi$. The dependence of the production rates on the complex
couplings is also obvious.
\vskip 10mm
\noindent
{\Large{\bf IV. Summary}}
\par
In this work we calculate the CP-violating effects to the process
$pp\rightarrow g g+X \rightarrow H^+H^-+X$ in the mSUGRA scenario at the
LHC collider, by taking into account the experimental bounds of electron
and neutron EDMs. The CP violation effects in this process are related to
the complex phases of $\mu$ and $A_{f}$ in the mSUGRA scenario. We choose
small CP phases of these parameters to neglect the mixing of neutral
Higgs bosons. In this case the CP effects to the process mainly come from
the complex couplings of Higgs-squark-squark. We find a strong dependence
of the production rates of the pair charged Higgs boson on the complex
couplings in the MSSM parameter space we chosen.

\vskip 5mm
\noindent{\large\bf Acknowledgement:}
This work was supported in part by the National Natural Science
Foundation of China(project numbers: 19875049), a grant from the
Education Ministry of China and the State Commission of Science and
Technology of China, and the Youth Science
Foundation of the University of Science and Technology of China.

\vskip 10mm
\begin{flushleft} {\bf Figure Captions} \end{flushleft}

{\bf Fig.1} The $\delta$ versus CP phases of $\mu$ and $A_{0}$
in mSUGRA with the parameters $m_0=100~GeV$, $m_{1/2}=150~GeV$,
$A=300~GeV$, $\tan\beta=4$. The solid curve is for $\Phi_{A_{0}}$ running from
zero to $0.1\pi$ while the phase of $\mu$ is set zero and the dashed curve
is for $\Phi_{\mu}$ running from zero to $0.1\pi$ while the phase of $A_{0}$
is set zero.

{\bf Fig.2} The $\delta$ versus CP phases of $\mu$ and $A_{0}$ in mSUGRA
with the parameters $m_0=100~GeV$, $m_{1/2}=150~GeV$,
$A=300~GeV$, $\tan\beta=30$. The solid curve is for $\Phi_{A_{0}}$ running from
zero to $0.1\pi$ while the phase of $\mu$ is set zero and the dashed curve
is for $\Phi_{\mu}$ running from zero to $0.1\pi$ while the phase of $A_{0}$
is set zero.

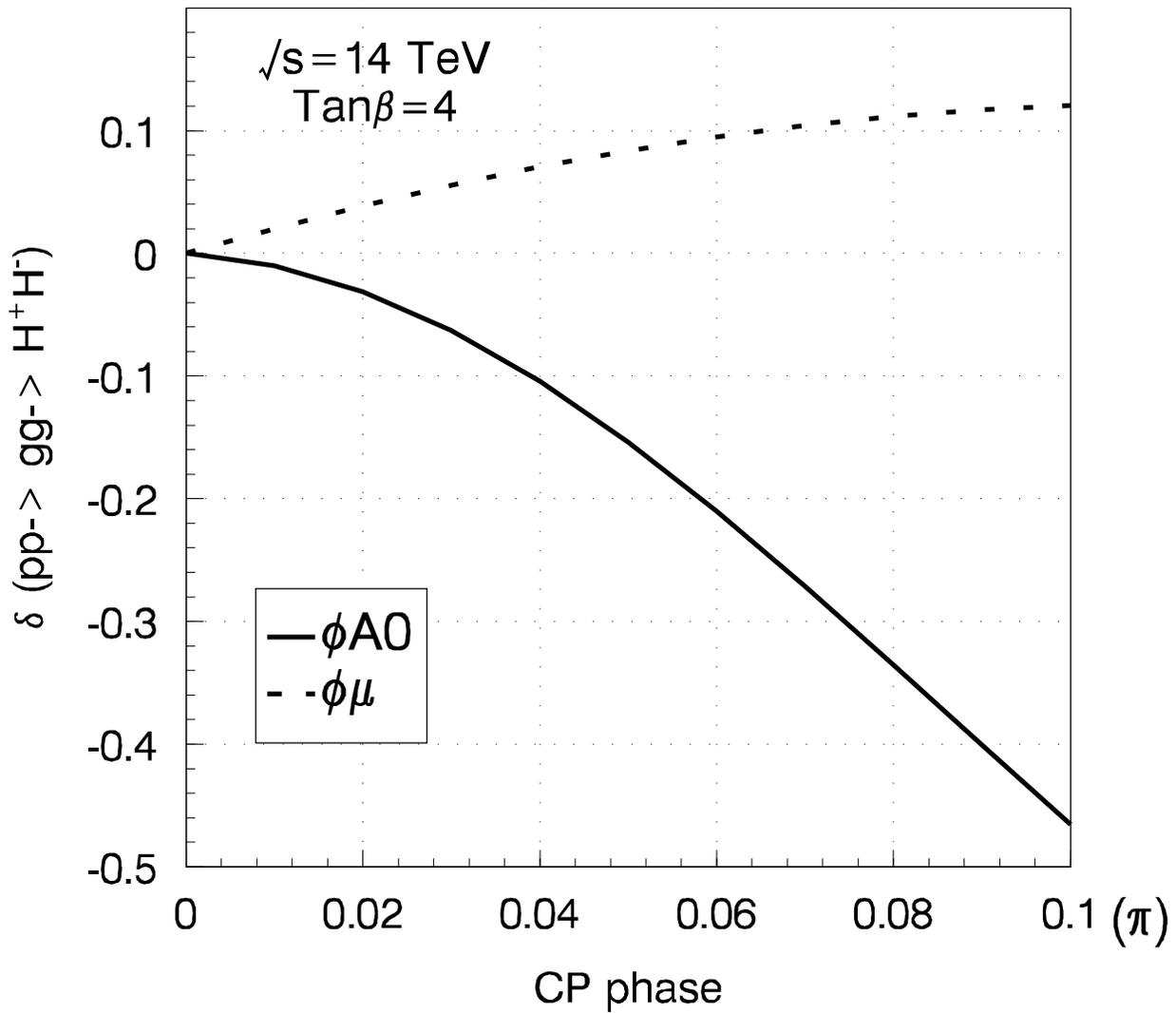
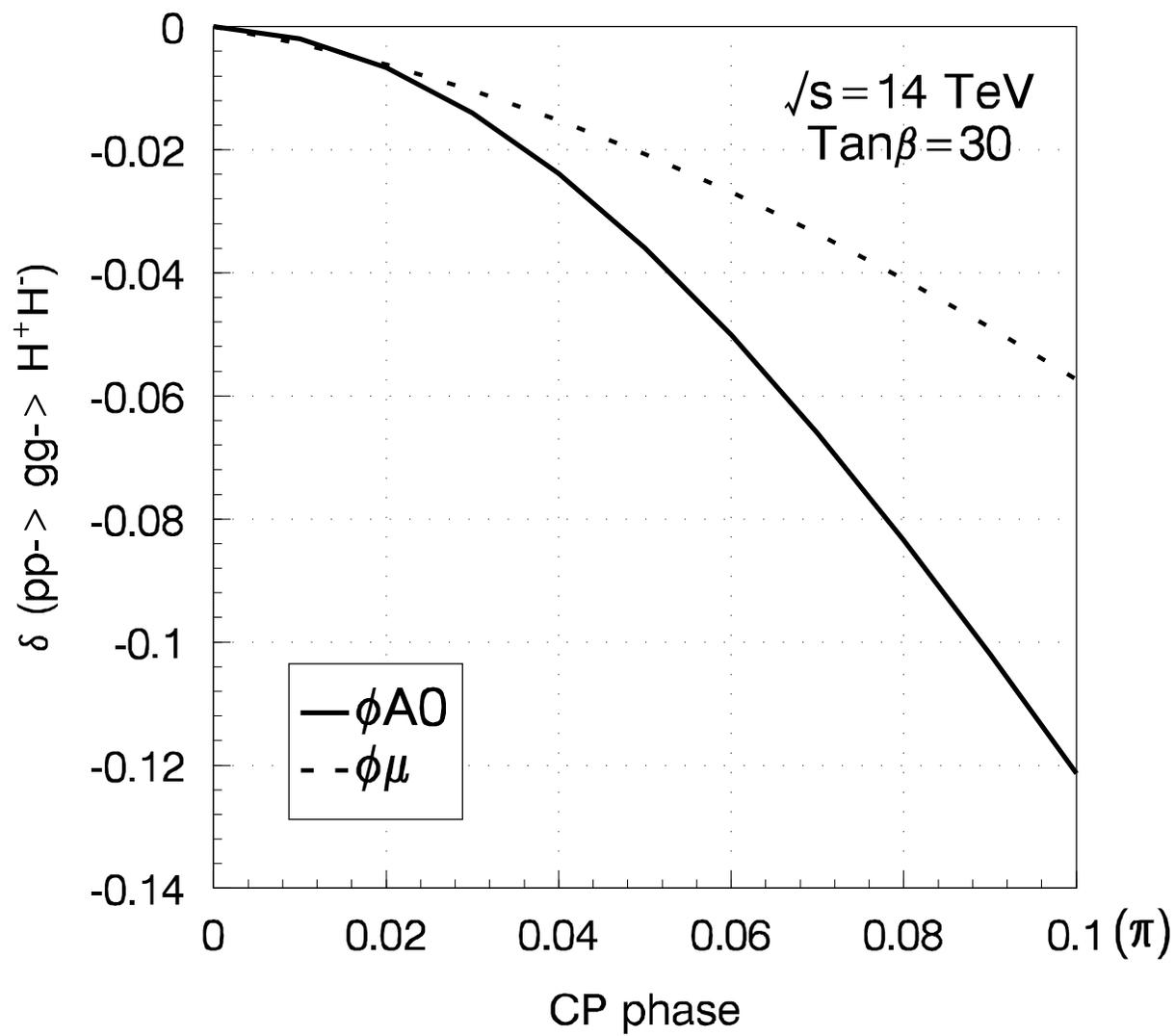
\end{document}